\documentclass[structabstract]{aa}  

\usepackage{natbib}	
\usepackage{txfonts,epsfig,graphicx,url,twoopt}
\usepackage[breaklinks=true]{hyperref} 
\hypersetup{colorlinks=true,citecolor=blue}
\bibpunct{(}{)}{;}{a}{}{,}   
\usepackage[usenames,dvipsnames]{color}

\begin{document}

\title{Ring shaped dust accumulation in transition disks }

   \author{P.~Pinilla\inst{1,2}, M.~Benisty\inst{3}, T.~Birnstiel\inst{4,5}}
   \institute{Universit\"at Heidelberg, Zentrum f\"ur Astronomie, Institut f\"ur Theoretische Astrophysik, Albert-Ueberle-Str. 2, 69120 Heidelberg, Germany\\
              \email{pinilla@uni-heidelberg.de}
        	       \and
	      Member of IMPRS for Astronomy \& Cosmic Physics at the University of Heidelberg
              \and
              Max-Planck-Institut f\"ur Astronomie, K\"onigstuhl 17, 69117 Heidelberg, Germany
              \and
              University Observatory Munich, Scheinerstr. 1, D-81679 M\"unchen, Germany
              \and
              Excellence Cluster Universe, Boltzmannstr. 2, D-85748 Garching, Germany}
   \date{Received  30 March 2012 / Accepted 26 July 2012}

 
\abstract
{Transition disks are believed to be the final stages of protoplanetary disks, during which a forming planetary system or photoevaporation processes open a gap in the inner disk, drastically changing the disk structure.  From theoretical arguments it is expected that dust growth, fragmentation and radial drift are strongly influenced by gas disk structure, and pressure bumps in disks have been suggested as key features that may allow grains to converge and grow efficiently.}
{We want to study how the presence of a large planet in a disk influences the growth and radial distribution of dust grains, and how
observable properties are linked to the mass of the planet.}
{We combine two-dimensional hydrodynamical disk simulations of disk-planet interactions with state-of-the-art coagulation/fragmentation models to simulate the evolution of dust in a disk which has a gap created by a massive planet.  We compute images at different wavelengths and illustrate our results using the example of the transition disk LkCa15.}
{The gap opened by a planet and the long-range interaction between the planet and the outer disk create a single large pressure bump outside the planetary orbit.  Millimeter-sized particles form and accumulate at the pressure maximum and naturally produce ring-shaped sub-millimeter emission that is long-lived because radial drift no longer depletes the large grain population of the disk.  For large planet masses around 9~$M_{\mathrm{Jup}}$, the pressure maximum and, therefore, the ring of millimeter particles is located at distances that can be more than twice the star-planet separation, creating a large spatial separation between the gas inner edge of the outer disk and the peak millimeter emission. Smaller grains do get closer to the gap and we predict how the surface brightness varies at different wavelengths.}
{}
  

\maketitle

\section{Introduction}     
\label{sec1}
Circumstellar disks are the birthsites of planets. The physical conditions and the evolution of these disks control the planet formation mechanisms. An important goal is to provide theoretical models and observational constraints to understand the various stages in the evolution of gas and dust in the disk. Decrease of mass accretion rate \citep{sicilia10,fedele10}, and near-infrared excess with time \citep{hernandez07, andrews11b} indicate that disks have a range of lifetimes from 1 to 10~Myr. 

With the advent of powerful infrared space telescopes such as \textit{Spitzer}, a new class of objects has been identified, called the transition disks \citep[e.g.,][]{espaillat10}.  Their spectral energy distribution (SED) and direct sub-millimeter (mm) imaging suggest that warm dust in the inner disk is strongly depleted compared to the outer disk. The small number of transition disks \citep{muzerolle10} suggests an inside-out evolution that occurs rapidly. Various mechanisms have been proposed so far to explain the inner disk clearing: photoevaporation winds \citep[e.g.][]{owen11}, grain growth \citep{klahr97, tanaka05} and dynamical interactions with companions \citep{lin79}. Transition disks are therefore excellent laboratories for planet formation models. 

The clearing of a gap by a companion or planet, from a simplistic point of view,  depends on the competition between the viscous torque from the disk and gravitational torques from the planet. For a laminar disk, a 1~$M_\mathrm{Jup}$ planet can clear a gap or hole. The recent discovery of a companion inside a massive disk in T Cha \citep{huelamo11} supports the scenario of a dynamical clearing, at least for this object.  However, models show that a single planet seems unlikely to be capable of creating the observed large holes, which require multiple systems \citep{zhu11, dodson11}. 

Interestingly, for most of these transition disks the inner cavity is not empty. They still present relatively high accretion rates ($\thicksim 10^{-8}M_{\odot}yr^{-1}$; see e.g.,\cite{calvet05, espaillat07}) which implies that the inner cavity is not completely empty and that some gas flows through the gap. To allow mass flowing, a limit for the planet mass can be inferred, depending on the disk viscosity \citep{lubow06}. In addition to the gas, some transition disks also present a strong near-infrared excess, indicating the presence of dust close to the star. \cite{rice06} studied the filtration of dust in the gap, considering a fixed size for the dust particles, and concluded that  increasing the planet mass from e.g. 0.5~$M_\mathrm{Jup}$ to 5~$M_\mathrm{Jup}$, the maximum particle size that sweeps into the gap decreases from $\sim10~\mu m$ to a few tenths of a micrometer.

Transition disks are potentially interesting laboratories to study processes related to the impact of planet formation on the disk. One of the most stubborn problems in planet formation is the so-called ``meter-size barrier". A one meter size object at 1~AU drifts towards the central star in timescales shorter than the growth timescales, impeding it to grow \citep[see e.g.,][]{brauer08, birnstiel10}. In addition, high relative velocities lead to numerous fragmentation collisions converting a large object into smaller dust particles. The same physical process happens to the millimeter-size particles that are observed in the outer regions of the disk \citep[e.g.,][]{wilner00, ricci10, guilloteau11}.  One possible solution to prevent the rapid inward drift and trap dust particles, is to consider pressure bumps \citep{klahr97, fromang05, brauer08, johansen09, pinilla11}. A  long-lived positive pressure gradient can lead dust particles to move outwards, causing an accumulation of dust at the location of the pressure maximum. A large pressure bump is expected in protoplanetary disks as a consequence of the presence of a massive planet in a disk. In fact, when a planet carves a gap in a disk, the gas surface density shows a significant depletion,  resulting in a large pressure bump at the outer edge of the gap. The dust material is trapped and piles up in this local pressure maximum where the gas motion is exactly Keplerian, and as a result there is no frictional drag between the gas and the particles. For that reason, not only do  the particles not drift anymore, they also do not experience the potentially damaging high-velocity collisions due to relative radial and azimuthal drift. Under those circumstances, growth to larger-than-usual sizes is expected, possibly even a breakthrough that leads to overcome the growth barrier. However, if turbulence is still strong enough, particles may fragment due to their relative turbulent velocities. This scenario is only possible if we assume the presence of a planet, formed by another mechanism than dust agglomeration.

In this paper, our goal is to test the idea that the outer edge of a planetary gap is a particle trap. We consider the dust evolution in a disk, where  the gas density profile is determined by its interaction with a massive planet. We then explore the case of LkCa~15, a transition disk \citep{mulders10}. The disk has been intensively observed at millimeter wavelengths \citep{pietu06, andrews11b} with a maximum angular resolution corresponding to 28~AU \citep{isella12}. The dust continuum images show a ring-like structure from $\sim$~42 to $\sim$120~AU, which is best fitted by a flat surface density profile. In addition, a  $\thicksim~$6~-~15$~M_{\mathrm{Jup}}$ planet in circular orbit and located at $15.7~\pm2.1$~AU, was claimed \citep{kraus11},  but not confirmed yet. 

We describe the numerical simulations of planet-disk interactions that are considered for this work, as well as the coagulation/fragmentation model of the dust evolution in Sect.~\ref{sec2}. Section~\ref{sec3} describes the results of the numerical simulations. In addition, we present observation predictions in Sect.~\ref{sect_obs}. Section~\ref{sec4} is a discussion of the main results of our work and Sect.~\ref{sec5} is a summary.

\section{Disk-planet interactions and dust evolution} 
\label{sec2}

For the disk-planet interaction process, we use the two-dimensional hydrodynamical code FARGO \citep{masset00}, that solves the Navier-Stokes and continuity equations. For the dust evolution, we use the code described in \cite{birnstiel10}, that computes the coagulation and fragmentation of dust grains, including turbulence, gas drag and radial drift.

\subsection{Codes}\label{sec2_1}

\begin{table}
\caption{FARGO parameters}    
\label{ref:table1}     
\centering                         
\begin{tabular}{c c }       
\hline\hline                 
Parameter & Values \\    
\hline                                            
   $M_{\star} [M_\odot]$ & $1.0$  \\   
   $\Sigma_0 [M_\star/r_p^2]$ (flared disk)& 1.26$\times 10^{-3}$  \\ 
   $\Sigma_0 [M_\star/r_p^2]$ (wedge disk)& 7.10$\times 10^{-4}$  \\  
   $M_P [M_{\mathrm{Jup}}]$ & $\{1.0, 9.0\}$\\
   $\alpha$ & $\{10^{-4}, 10^{-3}, 10^{-2} \}$  \\  
   $\beta$ & $\{-1, -0.5\}$\\
   $r_{\mathrm{in}}[r_p]$&$0.1$\\
   $r_{\mathrm{out}}[r_p]$&$7.0$\\
   $n_r\times n_\phi$&$[512\times1024]$\\
   $\epsilon [h(r)]$&$\{0.6, 0.8\}$\\
   Boundary&open\\
   conditions&\\
\hline                     
\end{tabular}  
\end{table}

FARGO uses finite differences and a fixed grid in cylindrical coordinates $(r, \phi)$.Its advantage lies in the calculation of the azimuthal advection. It allows each annulus of cells to rotate at its local Keplerian velocity and at the end of each timescale, the results are stitched together. This technique speeds up the corresponding calculations. FARGO has been used for numerous studies of disk evolution \citep[e.g.,][]{crida06, zhu11, regaly11}. We selected open inner and outer boundary conditions such that the material can leave the grid when the flow velocity at the inner or outer radial cells is pointing outwards. For this work, we only focus on the gas density profile at the outer edge of the open gap.

For the dust evolution, we consider that grains grow, crater and fragment due to radial drift, turbulent mixing and gas drag. Brownian motion, turbulence, vertical settling, radial and azimuthal drifts are taken into account to calculate the relative velocities of dust particles. For micron-sized particles, the relative velocities mainly result from Brownian motion and settling. These grains grow by coagulation  as a result of  van der Waals interactions. As they grow, they start to decouple from the gas, and turbulent motion as well as radial drift become the main sources for relative velocities. 

When dust grains reach  sizes with high enough velocities, they encounter destructive collisions. Particles above the fragmentation velocity always cause fragmentation if the impactor mass is within one order of magnitude of the target, otherwise they cause erosion. The transition from the 100\% sticking probability to 100\% fragmentation/erosion probability is linearly increasing between $0.2~\times~ v_f$ to $v_f$, as in \cite{birnstiel10}.

The fragmentation velocities of the dust grains can be estimated based on  laboratory experiments and theoretical work of collisions for silicates and ices \citep[e .g.,][]{blum08, schafer07, wada09}. For silicates, they are of the order of few $\mathrm{m~s}^{-1}$ \citep[e .g.,][]{blum08}, and increase  with the presence of ices  \citep{schafer07, wada09}. \\

\begin{table}
\caption{Dust model parameters}    
\label{ref:table2}     
\centering                         
\begin{tabular}{c c }       
\hline\hline                 
Parameter & Values \\    
\hline 
   $M_{\star} [M_\odot]$ & $1.0$  \\   
   $M_{\mathrm{disk}} [M_{\mathrm{Jup}}]$ & $55.0$  \\   
   $\alpha$ & $\{10^{-3}, 10^{-2} \}$  \\  
   $r_{\mathrm{in}}[\mathrm{AU}]$&$2.0$\\
   $r_{\mathrm{out}}[\mathrm{AU}]$&$140.0$\\                             
   $R_{\star}$ [R$_\odot$] &$1.7$\\
   $T_{\star}$ [K]&$4730$\\
   $\rho_s$ [g/cm$^3$]& $1.2$\\
   $v_f$ [cm/s]& $1000$\\
\hline                     
\end{tabular}  
\end{table}

To characterize how well the dust particles couple to the gas, we refer to the Stokes number St, which is defined as the ratio between the turn-over time of the largest eddy ($1/\Omega$) and the stopping time of the particle due to the friction with the gas. In the presence of a drag force, big particles are not affected by the drag and they move with their own Keplerian velocity, implying that $\mathrm{St}~\gg~1$. When $\mathrm{St}~\ll~1$, the particles are small enough to be well coupled to the gas and  move with the gas. 

In the Epstein regime, where the ratio between the mean free path of the gas molecules $\lambda_{\mathrm{mfp}}$ and the sizes of the particles $a$ is $\lambda_{\mathrm{mfp}}/a~\geq~4/9$, the Stokes number is at the midplane given by \cite{birnstiel10},

\begin{equation}
	\textrm{St}=\frac{a\rho_s}{\Sigma_g}\frac{\pi}{2},
  	\label{eq1}
\end{equation}

\noindent where $\rho_s$ is the intrinsic density of the dust particles, and $\Sigma_g$ is the gas surface density given by $\Sigma_g=\int^{\infty}_{\infty}\rho(r,z) dz$, where $\rho(r,z)$ is the total gas density.

Due to the sub-Keplerian velocity of the gas, particles that are large enough and do not move with the gas velocity experience a head wind, leading them to lose angular momentum and move inwards. The radial component of the dust velocity is the sum of a term corresponding to the frictional drag force (which opposes the motion of the dust grain) and of a term corresponding to the drift velocity, $u_{\mathrm{drift}}$, with respect to the gas, which directly depends on the radial pressure gradient $\partial_r P$. It is given by:

\begin{equation}
	u_{\mathrm{r,dust}}=\frac{u_{\mathrm{r,gas}}}{1+\textrm{St}^2}+\frac{1}{\textrm{St}^{-1}+\textrm{St}} \frac{\partial_r P}{\rho \Omega},
  	\label{eq2} 
\end{equation}

\noindent where $u_{\mathrm{r,gas}}$ is the gas radial velocity. The first term is the dominant contribution for St~$\ll~1$.  Equation \ref{eq2} assumes low values for the dust-to-gas ratio. The drift velocity is usually  negative, unless the pressure gradient is positive at a given radius. Particles with St~=~1 experience the highest possible radial drift.  

When particles reach certain sizes due to coagulation, they have relative velocities that may be high enough to cause fragmentation. For particles with $\mathrm{St}~\sim~1$,  the maximum turbulent relative velocity is given by \citep{ormel07}

\begin{equation}
	\Delta u_{max}^2\simeq \frac{3}{2}\alpha  \mathrm{St} c_s^2,
  \label{eq3}
\end{equation}

\noindent where $\alpha$ is the turbulence parameter \citep{shakura73}. 

For dust particles with $\alpha/2~\lesssim~\mathrm{St}\lesssim~1$ relative velocities are dominated by turbulent motion and radial drift. However, if the radial drift is reduced due to a positive pressure gradient, an approximation of the maximum size of the particles is obtained when the turbulent relative velocities are as high as the fragmentation velocity $v_f$, therefore:

\begin{equation}
	a_{\mathrm{max}}\simeq\frac{4\Sigma_g}{3\pi \alpha\rho_s} \frac{v_f^2}{c_s^2},
	  \label{eq4}
\end{equation}
 
\noindent  The dust grain distribution $n(r,z,a)$ depends on coagulation and fragmentation collisions and is described by the Smoluchowski equation \citep[see][Eq.~35 and Eq.~36]{birnstiel10}. The vertically integrated dust surface density distribution per logarithmic bin is

\begin{equation}
	\sigma (r,a)=\int_{-\infty}^{\infty} n(r,z,a)\cdot m\cdot a dz
  \label{eq5}
\end{equation}

\noindent where $m$ is the mass of a single particle of size $a$, hence the total dust surface density is:

\begin{equation}
	\Sigma_d(r)=\int_0^\infty \sigma (r,a)d\ln a.
  \label{eq6}
\end{equation}

For the coagulation process, the midplane is the most important region, therefore the vertically integrated dust surface density distribution is a good approximation for describing the dust grain distribution. The advection-diffusion differential equation that describes the evolution of the dust surface density $\Sigma_d$ for a single dust size, can be written in cylindrical coordinates as:

\begin{equation}
	\frac{\partial \Sigma_d}{\partial t} + \frac{1}{r}\frac{\partial}{\partial r}\left( r \Sigma_d u_r\right)-\frac{1}{r}\frac{\partial}{\partial r} \left(r \Sigma_g D_d \frac{\partial }{\partial r}\left[\frac{\Sigma_d}{\Sigma_g}\right]\right)=0,
  \label{timeevo}
\end{equation}

where $D_d$ is the dust diffusivity  and  is equal to $D_d~=~\nu/(1+\mathrm{St}^2)$, if the gas diffusivity is taken  to be the turbulent gas viscosity $\nu~=~\alpha~c_s~h$. Equation \ref{timeevo} is solved for each size using the flux-conserving donor-cell scheme \citep[see][Appendix A]{birnstiel10}.\\

For a massive planet on a fixed orbit, the gap opening process reaches a quasi-steady state in timescales that are  shorter ($\lesssim~1000$ orbits) than the million-years timescales of the entire process of dust evolution. For this reason,  we use the gas surface density after one thousand orbits ($\sim 10^{-2}$~Myr) and take it as an input for the dust evolution models. Therefore, we assume that the gas surface density remains stationary from one thousand orbits  to several million years.

We consider that after the giant planet is formed, the disk remains dust-rich and we study how the remaining dust  evolves after the gap is open and has reached a quasi-steady state.  We consider that the gas does not have any feedback from the dust and stays static during the dust evolution process since the dust density is always less than the gas density.  In addition, since the orbital timescales are much shorter than the dust evolution timescales and there are no strong asymmetries produced by the planet, we azimuthally average the surface density for our models. With these assumptions, the dust evolution at the outer edge of the gap is computed in an accurate way.

 \begin{figure*}
   \centering
   \includegraphics[width=18.0cm]{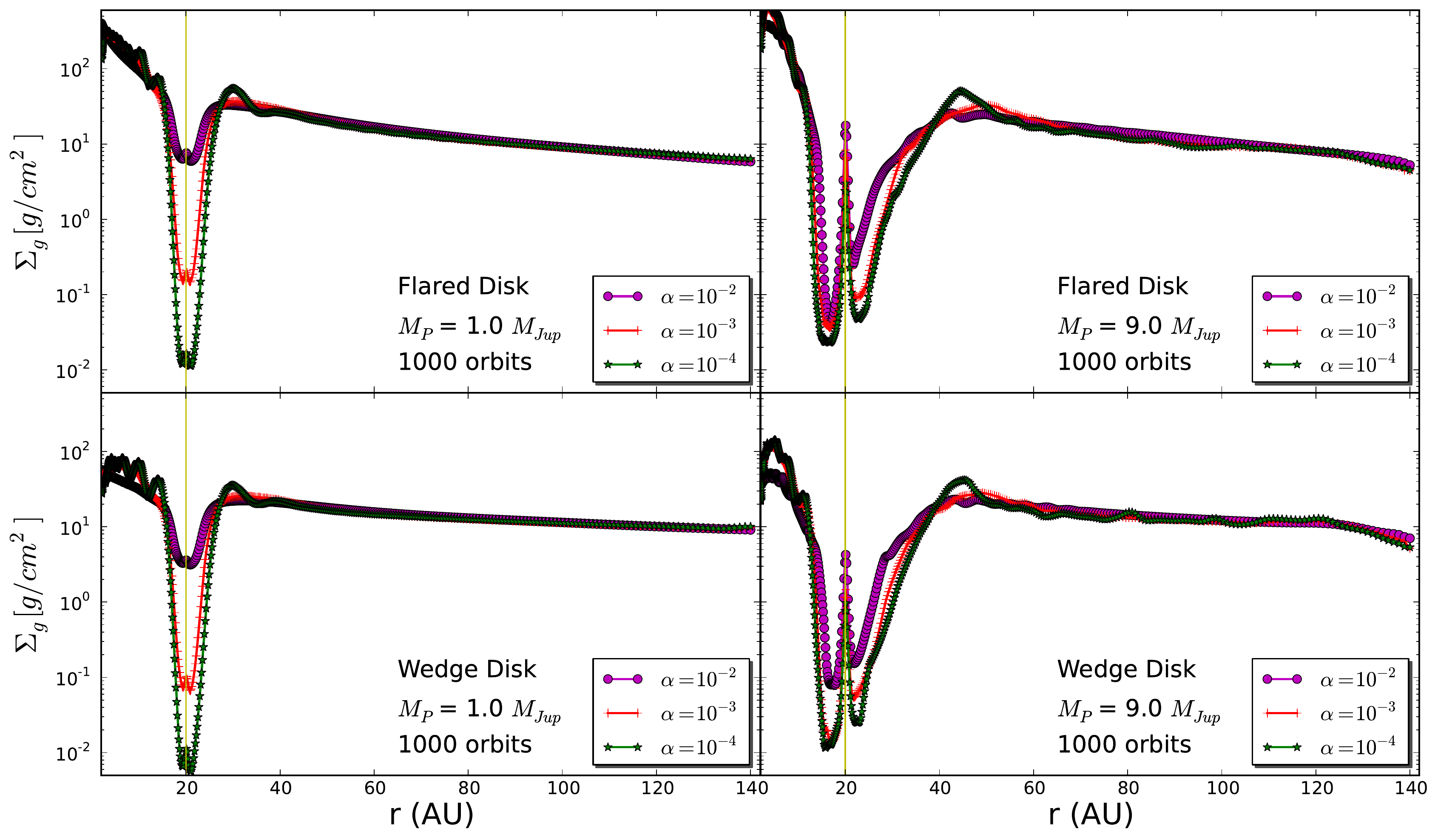}
   \caption{Azimuthally averaged surface density  after 1000 orbits and three different values of $\alpha$-turbulence for a gap created by a planet located in a fixed orbit at 20~AU of 1~$M_\mathrm{Jup}$ in a flared disk (top-left), 9~$M_\mathrm{Jup}$ in a flared disk (top-right), 1~$M_\mathrm{Jup}$ in a wedge disk (bottom-left) and 9~$M_\mathrm{Jup}$ in a wedge disk (bottom-right). The peak  of the gas surface density at the position of the planet is due to the strong gravitational effect of the massive planet on the gas of the disk.} 
   \label{Fig1}
\end{figure*}

\subsection{Set-up} \label{sec2_2}

For our simulations, we assume a locally isothermal disk and two different radial density profiles $\Sigma~\propto~r^{\beta}$. When the disk scale increases with radius as $h/r \propto r^{1/4}$ (flared disk) as suggested by modelling of T Tauri disks \citep{dalessio01}, the temperature scales as $T \propto r^{-1/2}$, and $\beta=-1$. When $h/r$ is constant (wedge disk), the temperature scales as $T \propto r^{-1}$ and $\beta=-0.5.$ The $\alpha$ turbulence parameter is considered to have values between $10^{-4}$ to $10^{-2}$, in agreement with magnetorotational instability (MRI) of active disks \citep{johansen05}.  The numerical simulations are performed with $n_r~=~512$ radial logarithmically, and $n_\phi~=~1024$ azimuthal uniformly distributed grid cells. 

All parameters such as the stellar properties, kinematic viscosity, disk mass, the scale height and the gas density profile of the disk are taken to be the same between the hydrodynamical gas evolution and the dust coagulation simulations.

For FARGO simulations, we use normalized units such that $G~=~M_{\star}+M_P=1$ and the location of the planet is at $r~=~1$. The simulations are performed from $r_{\mathrm{in}}~=~0.1$ to $r_{\mathrm{out}}~=~7.0$, such that the planet is at $r_p=$~20~AU for a grid from 2~AU to 140~AU. The initial gas surface density $\Sigma_0$ at the position of the planet is fixed $\Sigma=\Sigma_0(r/r_p)^{\beta}$, such that it is around $\sim$~1-1.5 the minimum mass solar nebula \citep{hayashi81}, and the disk mass is around $\sim$0.055 $M_{\star}$. We consider a planet at a fixed orbit, with two different masses, 1~$M_{\mathrm{Jup}}$ and 9~$M_{\mathrm{Jup}}$. The planet is introduced slowly into the smooth disk to avoid numerical issues. Planetary accretion is not taken into account since this process is still debated.  We use $M_{\star}~=~1~M_\odot$, and consequently, $M_{\mathrm{disk}}~=~55~M_{\mathrm{Jup}}$. 
Finally, the gravitational effect of the planet is smoothed out, such that the gravitational potential $\phi$  is softened over distances comparable to the disk scale height:

\begin{equation}
	\phi=-\frac{GM_{P}}{(r^2+\epsilon^2)^{\frac{1}{2}}},
  \label{eq7}
\end{equation}

\noindent where  $\epsilon$ is taken to be $0.6~h$. The main FARGO parameters are summarized in Table~\ref{ref:table1}.\\

For the dust evolution,  we consider that all particles are initially 1~$\mu m$ large, and a dust-to-gas ratio is  0.01, with an intrinsic volume density of $\rho_s = 1.2~\mathrm{gcm}^{-3}$ and a fragmentation velocity of $v_f~=~1000\mathrm{cm~s}^{-1}$.  Two additional stellar parameters are taken into account, the stellar radius which is taken to be $R_{\star}~=~1.7~R_\odot$ and the star effective temperature $T_{\star}=4730$~K, as typical values for T Tauri stars. The parameters of the  dust evolution simulations are  in Table~\ref{ref:table2}.

\section{Results} \label{sec3}


%
\subsection{Gas density profile}  \label{sec3_1_1}

Figure~\ref{Fig1} shows the azimuthally averaged gas surface density after 1000 orbits in four different cases: a 1~$M_\mathrm{Jup}$ and a 9~$M_\mathrm{Jup}$ planet on a fixed orbit at 20~AU in a flared and wedge disk respectively. In each case, three different values for $\alpha$ turbulence parameter are considered. 

 \begin{figure*}
   \centering
   \includegraphics[width=18.0cm]{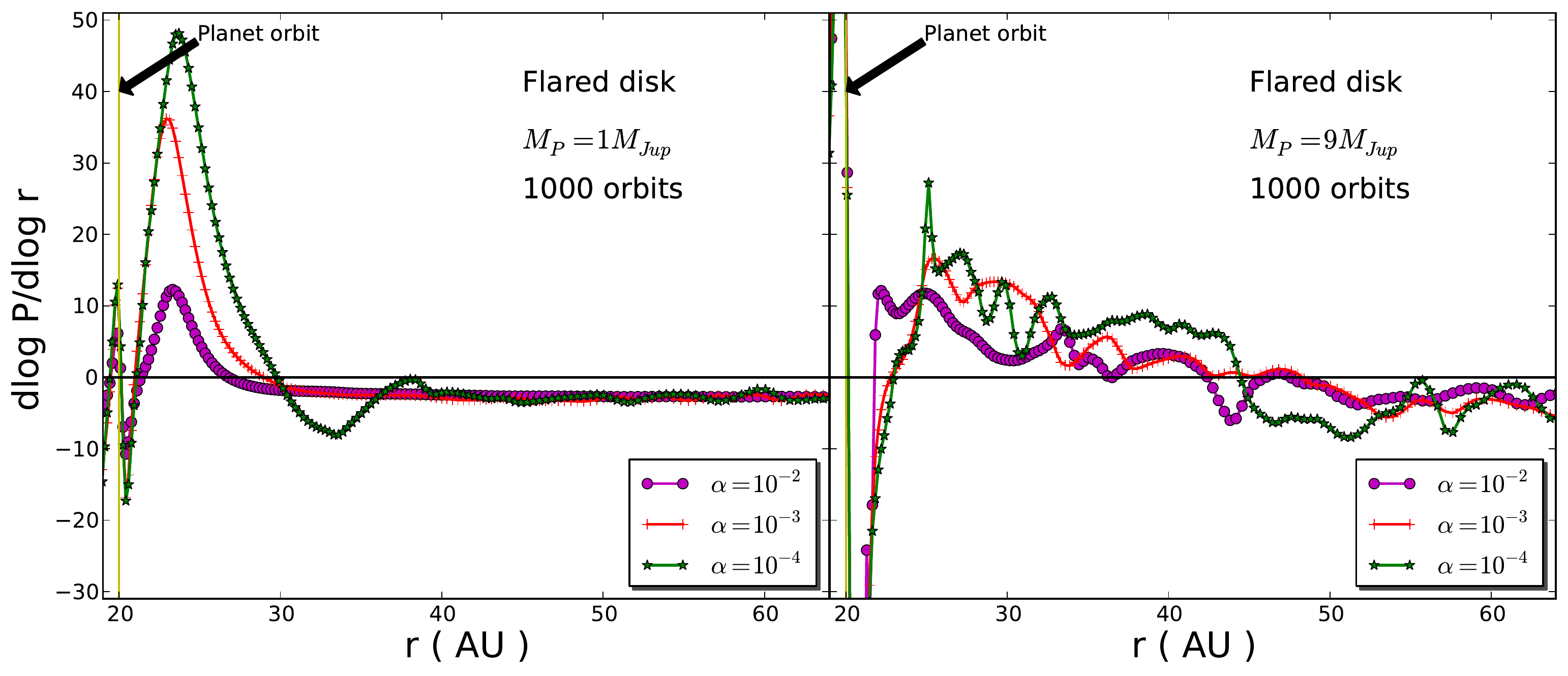}
   \caption{Radial pressure gradient after 1000 orbits and three different values of $\alpha$ turbulence for a gap created by a planet located in a fixed orbit at 20~AU of 1~$M_\mathrm{Jup}$ and 9~$M_\mathrm{Jup}$ in a flared disk. }
   \label{Fig2}
\end{figure*}

With a  1~$M_\mathrm{Jup}$ planet and a high value of $\alpha~=~10^{-2}$, the viscous torque on the gas exceeds the gravitational torque by the planet, resulting in a much less pronounced gap than in the cases of $\alpha~=~10^{-3}$ and $\alpha~=~10^{-4}$. The gap is replenished easily because of the high viscosity of the disk as already predicted by different authors \citep[e.g][]{devalborro06, crida06}. However, for all $\alpha$ values, the width of the gap is roughly the same. It is independent of the viscosity and it depends basically on the mass and location of the planet. We notice that with low viscosity, the density waves that are produced by the planet-disk interaction at the gap edges, become more pronounced but they eventually damp with radius. Indeed, with a 1~$M_\mathrm{Jup}$ planet, these waves disappear beyond 40~AU and the disk surface density profile follows the initial power law. In this case, the main difference between the two disk geometries that we consider, is that for each value of $\alpha$, the gap is slightly deeper for a wedge than for a flared disk. 

 \begin{figure}
   \centering
   \includegraphics[width=9.0cm]{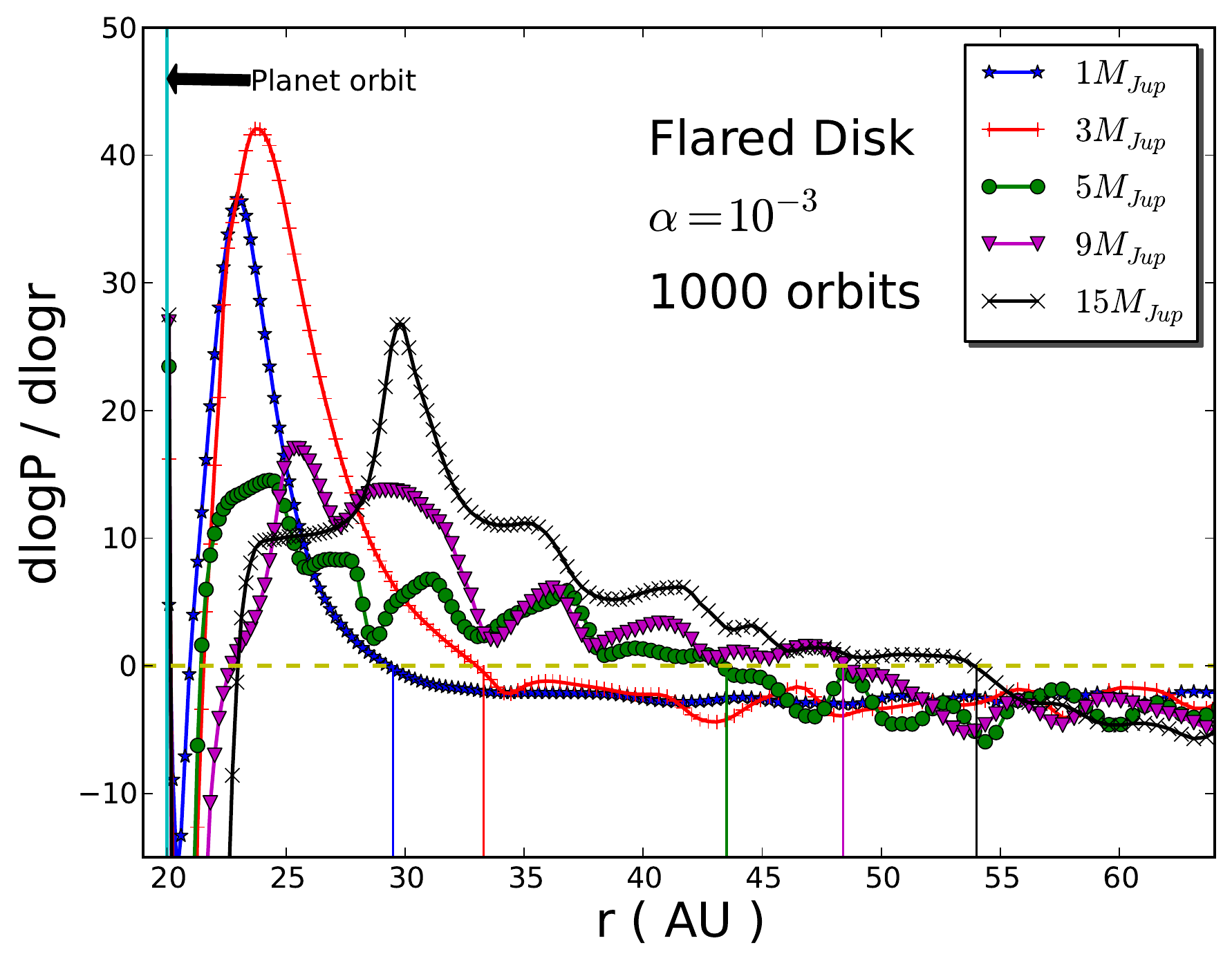}
   \caption{Effect of the planet mass on the radial pressure gradient after $\sim$~1000 orbits in a flared disk,  $\alpha~=~10^{-3}$ and five different values of the planet mass. The planet is located at 20~AU. The vertical lines indicate the locations of the pressure maximum.}
   \label{gradient}
\end{figure}

With a 9~$M_\mathrm{Jup}$ planet, the gravitational torque dominates the viscous torque and the depth of the gap is almost independent  of the values that we consider for $\alpha$. After 1000 orbits the surface density reaches almost the same values near the planet and the width of the gap slightly changes for the different values of turbulence, when the disk is less turbulent the gap is moderately wider. The density perturbation waves are more evident in this situation than for 1~$M_\mathrm{Jup}$, however, for $\alpha~=~10^{-3}-10^{-2}$, they again damp with radius, and at  around 60~AU the gas surface density follows the initial power law. After 1000 orbits and $\alpha~=~10^{-4}$, the density perturbation due to the planet is still propagating, for this reason the wiggles in the density profile are stronger than in the cases of $\alpha~=~10^{-3}-10^{-2}$. These wiggles are still present for $r~>~100$~AU and they change slightly  with the geometry of the disk.

In all four cases, it is possible to distinguish a peak in the surface density at the position of the planet, due to circumplanetary material. This is a natural consequence of the planet insertion in a smooth disk, and because the accretion onto the planet is not considered,  the peak remains unaltered during the simulations.

\subsection{Pressure gradient}  \label{sec3_1_2}

As explained before, the rapid inward drift of the particles stops if the gradient of pressure is positive and the particles pile up in the pressure maxima. Depending on the strength of the pressure gradient, the particles can be trapped or not  in the pressure bump \citep{pinilla11}.  Figure~\ref{Fig2} shows the radial pressure gradient considering the azimuthally averaged gas surface density after one thousand orbits, in the case of a flared disk and with an isothermal equation of state. The pressure is given by $P(r)~=~\rho(r)~c_s^2(r)$, where the sound speed is defined as

\begin{equation}
	c_s=\sqrt{\frac{k_B T}{\mu m_p}},
  \label{soundspeed}
\end{equation}

 \begin{figure*}
   \centering
   \includegraphics[width=18.0cm]{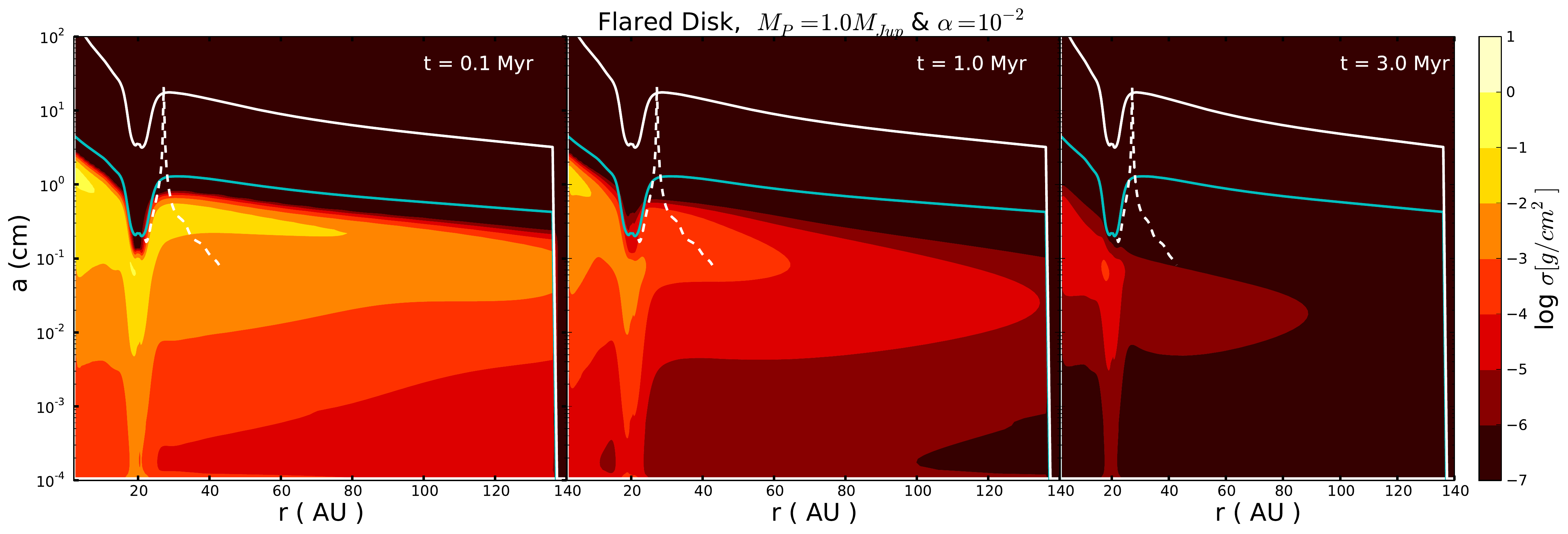}\\
   \includegraphics[width=18.0cm]{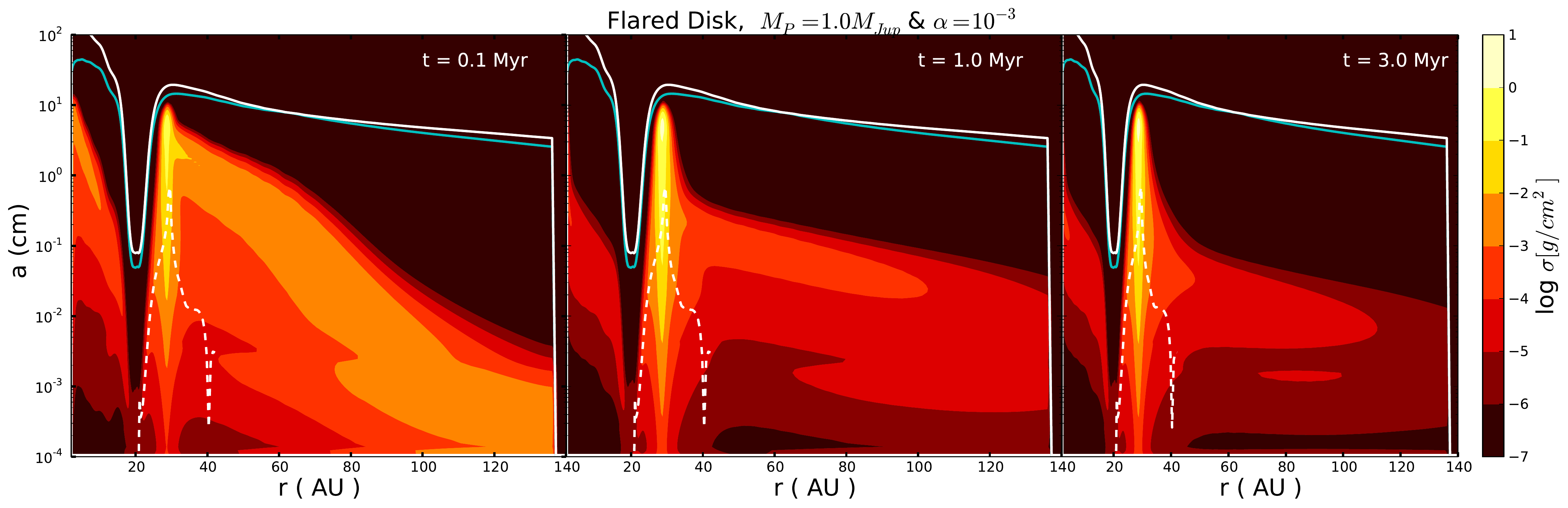}\\
   \includegraphics[width=18.0cm]{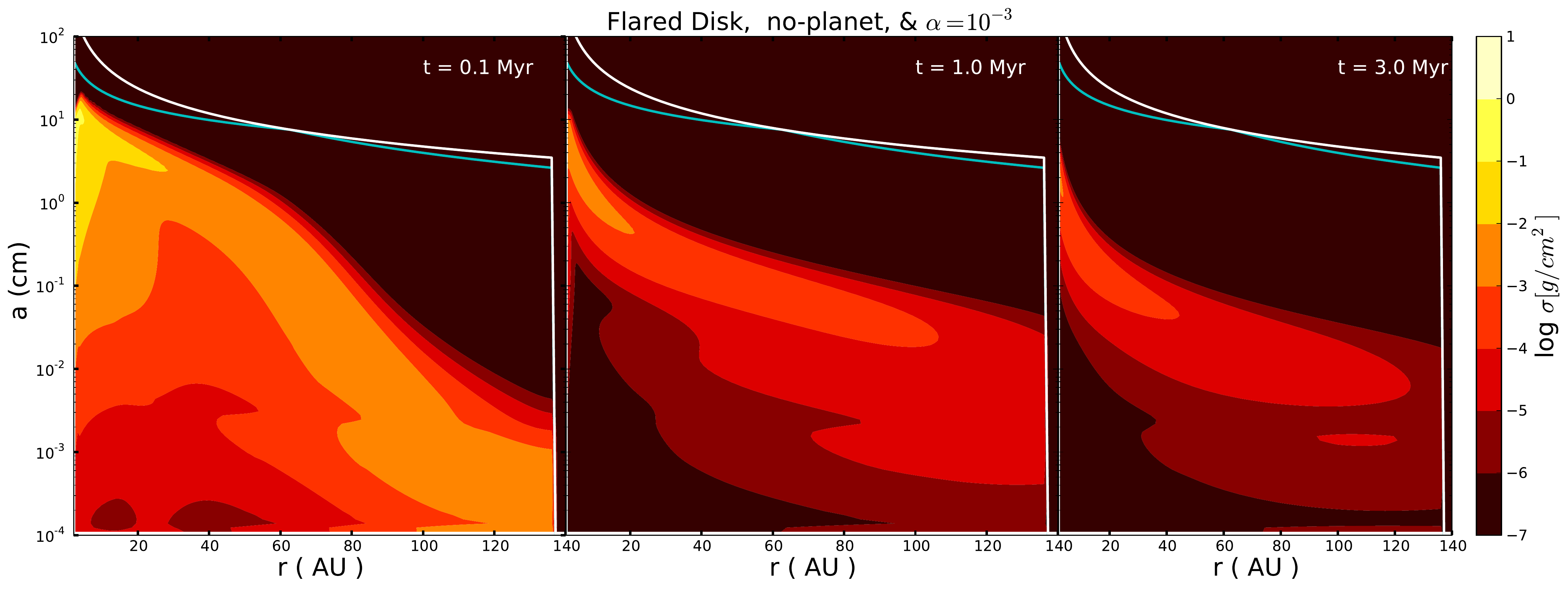}\\
   \caption{Vertically integrated dust density distribution at different times of evolution. Case of a surface density with a gap created by 1~$M_\mathrm{Jup}$ located in a fixed orbit at 20~AU in a flared disk and $\alpha~=~10^{-2}$ (top panel), $\alpha~=~10^{-3}$ (middle panel). Case with no planet and $\alpha~=~10^{-3}$  (bottom panel). The solid white line represents the particle size corresponding to St~=~1, while the blue line corresponds to the fragmentation barrier according to Eq.~\ref{eq4} \citep{brauer08, birnstiel10}. The dashed line corresponds to the size for which particles drift faster than those dragged by the gas. Particles above the dashed line are perfectly trapped.}
   \label{Fig3}
\end{figure*}

where  $k_B$ is the Boltzmann constant, $m_p$ is the proton mass and $\mu$ is the mean molecular mass, which is $\mu=2.3$ in proton mass units. If the temperature is considered as a smooth decreasing function with radius, density variations induce pressure inhomogeneities. We can see that for a 1~$M_\mathrm{Jup}$ planet, there is a noticeable difference in the amplitude of the pressure gradients for $\alpha~=~10^{-2}$ and $\alpha~=~10^{-4}-10^{-3}$. The pressure gradient is positive from $\sim$22~AU to $\sim$30~AU, and since $u_\mathrm{drift}~\propto~\partial_r P$ (Eq.~\ref{eq2}), the dust particles have positive drift velocities and move outwards. In all cases, the zero point of the pressure gradient  is located at $\sim$~10~AU (with a slight variation with viscosity) from the planet orbit and does not change with the geometry of the disk, since the density  profile behaves almost the same for flared and wedge disks. \\

For a 9~$M_\mathrm{Jup}$ planet orbiting at 20~AU, the pressure gradient on average behaves similarly for each of the $\alpha$ values. The wiggly profile of the gradient is due to the density waves that are produced by the interaction between the planet and the disk. For this planet mass and all values of $\alpha$, the pressure gradient is positive right after the planet location and remains positive until around 48~AU, and  then becomes again negative for both cases of flared and wedge disks. The influence of the planet mass on the range where the pressure gradient is positive is remarkable.

In both mass regimes, it is clear that the location of the pressure maximum differs from the outer edge radius of the gap, where the gas density again starts to increase. Therefore, if the pressure gradient is high enough to moderate the rapid inward drift, dust particles are trapped further away from the gap outer edge in the gaseous disk. This implies that the gap width for the gas is smaller than the one for the dust.

 \begin{figure*}
   \centering
   \includegraphics[width=18.0cm]{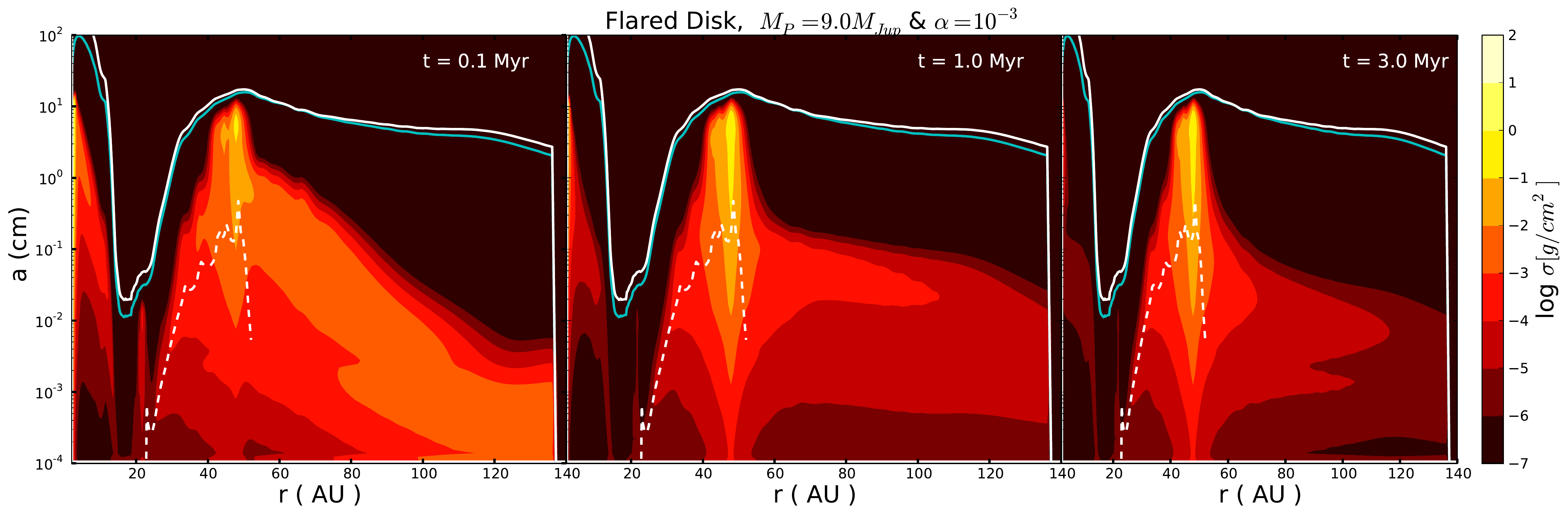}
   \caption{Vertically integrated dust density distribution at different times of evolution for a surface density with a gap created by 9~$M_\mathrm{Jup}$ located in a fixed orbit at 20~AU in a flared disk and $\alpha~=~10^{-3}$. Lines are as in Fig.~\ref{Fig3}.}
   \label{Fig4}
\end{figure*}

Figure~\ref{gradient} summarizes these findings, and shows pressure gradient curves for additional planet masses, $M_p~=~\{1,3,5,9,15\}M_{\mathrm{Jup}}$ with orbital radius $r_p~=~20$~AU. The pressure gradient has its zero point at around $\{29.5,33.5,43.5,48.5,54.0\}$~AU respectively. 

As a comparison, we use the Hill radius $r_H$, defined as $r_H~=~r_p~(M_p/3M_\star)^{1/3}$. In spite of the fact that the outer gap edge in the gas is at most 5~$r_H$ \citep[see e.g.][]{dodson11}, we find that the location where dust accumulates  (i.e., the location of the maximum of the pressure) is at $\sim7~r_H$ for 1 and 3~$M_{\mathrm{Jup}}$ planets, and $\sim10~r_H$ for  planets with $M_p\geq5~M_\mathrm{Jup}$. This result will be discussed in more detail in Sect.~\ref{sec4}. 

Note that for the FARGO simulations in the case of a 15~$M_{\mathrm{Jup}}$ planet, we insert the planet in the disk  after more orbits than in the other cases studied in Sect.~\ref{sec3_1_1}, we use non-reflecting boundary conditions and a smoothing parameter of $\epsilon~=~0.8h$ (see Eq. \ref{eq7}) to smooth out the strong density waves produced in this scenario.

\subsection{Dust size evolution}  \label{sec3_1_3}
%

 \begin{figure}
   \centering
   \includegraphics[width=9.0cm]{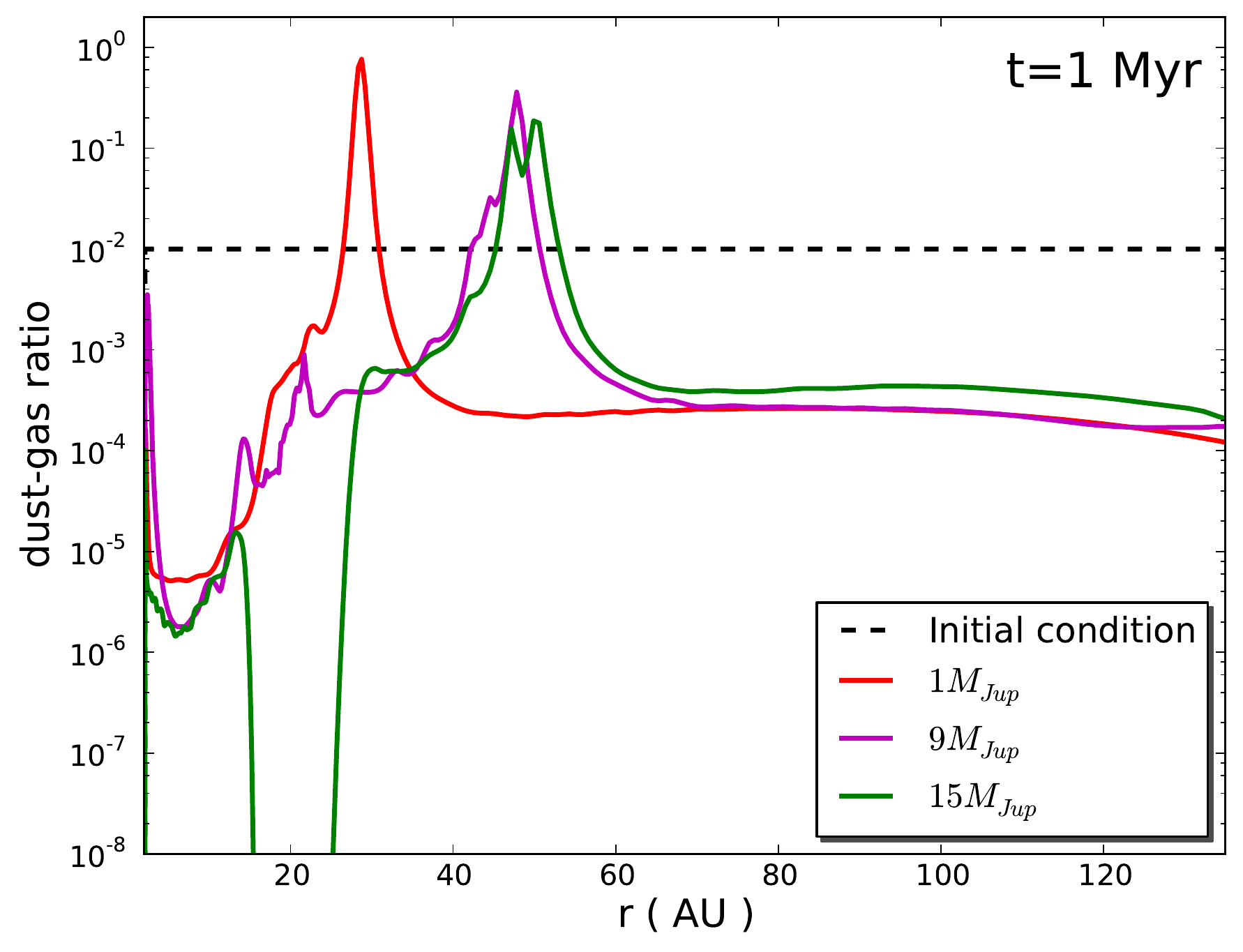}
   \caption{Dust-to-gas ratio for 1, 9 and 15~$M_{\mathrm{Jup}}$ planet, after 1~Myr of dust evolution}
   \label{Fig_ratio}
\end{figure}

For the study of the density distribution of dust particles, we again focus on the flared disks, since this disk geometry is in good agreement with observations of T Tauri disks \citep{dalessio01}.  For this section, the surface density peak at the planet location was smoothed out to avoid artificial effects at this location. 

Top and middle panels of Fig.~\ref{Fig3} show the vertically integrated dust density distribution at different evolution times for a gas surface density with a  gap created by a 1~$M_\mathrm{Jup}$ planet located in a fixed orbit at 20~AU, and two different values of $\alpha$ turbulence. As explained before the timescales of the gap opening process are much shorter than the whole dust evolution process, which allows us to take the gas surface density  after $\sim$~1000 orbits and  consider that it remains quasi-stable during the dust evolution. The solid white line has the shape of the gas surface density profile and is calculated when the particle size corresponds to a Stokes number of unity (Eq.~\ref{eq1}). The blue line is the maximum size that particles can reach before they have velocities higher than the fragmentation limit. It is important to note that the maximum size of the particles is inversely proportional to the turbulence parameter (Eq.~\ref{eq4}), for this reason the maximum possible grain size (blue line) differs by one order of magnitude for each $\alpha$ turbulent parameter considered.

The dashed line provides an approximation of the minimum size of the particles that are trapped in the pressure maximum. Particles with sizes over the dashed line are trapped and the ones below the line are dragged by the gas.  This condition is found when the radial component of the dust velocity  (Eq.~\ref{eq2}) is positive, therefore:

\begin{equation}
	\textrm{St}>-\frac{u_{\mathrm{r,gas}}}{\partial_r P} \rho \Omega,
  \label{cond1}
\end{equation}

which in terms of the particle size and pressure gradient can be written as:

\begin{equation}
	a_{\mathrm{critical}}=\frac{6\alpha\Sigma_g }{\rho_s \pi |(d\log P / d\log r)|} \left|\left(\frac{3}{2}+\frac{d\log \Sigma_g}{d\log r}\right)\right|.
  \label{cond2}
\end{equation}

\subsubsection{Effect of turbulence}
We notice a significant influence of the trapping of dust particles on the strength of the pressure gradient. Micron-size particles (St~$\ll 1$), which are easily mixed due to turbulence, are more difficult to trap, they go through the gap and finally drift towards the star. Dust grains with larger Stokes number can perfectly be trapped because they feel a tail-wind due to the fact that the gas moves with super-Keplerian velocity where the pressure gradient is positive, resulting in an outwards movement for those particles.

For $\alpha~=~10^{-2}$, the pressure gradient is less steep, the turbulent mixing and drag along the accretion flow wins over the trapping, and the dust gets lost onto the star. Even when the pressure gradient is positive in a range ($\sim$~22-30)~AU, it is not sufficient to trap the particles in the pressure maximum. Indeed, after 3~Myr there is almost no dust remaining in the whole disk.

 \begin{figure*}
  \centering
   \begin{tabular}{cc}
   \includegraphics[width=8.5cm]{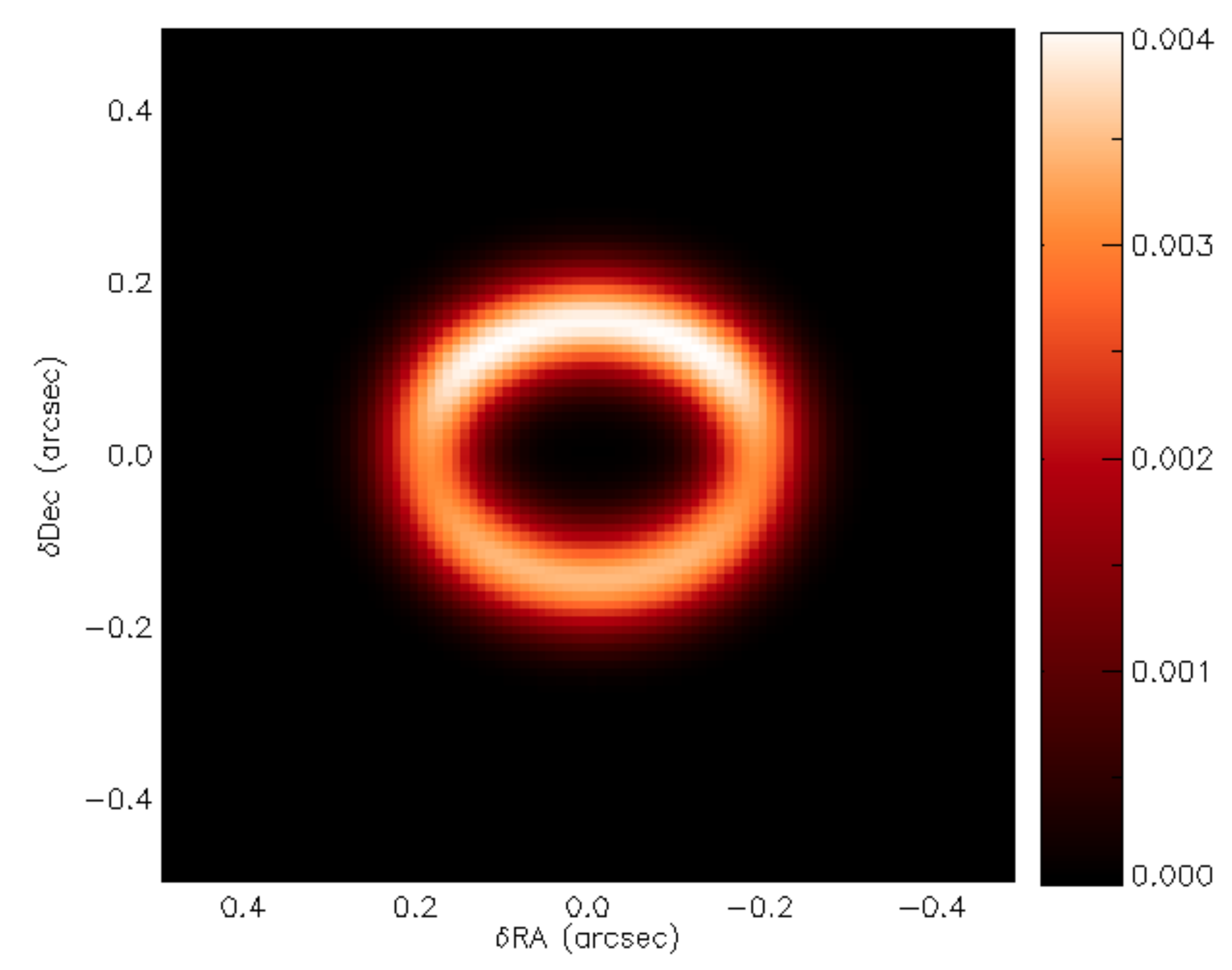} & 
      \includegraphics[width=8.5cm]{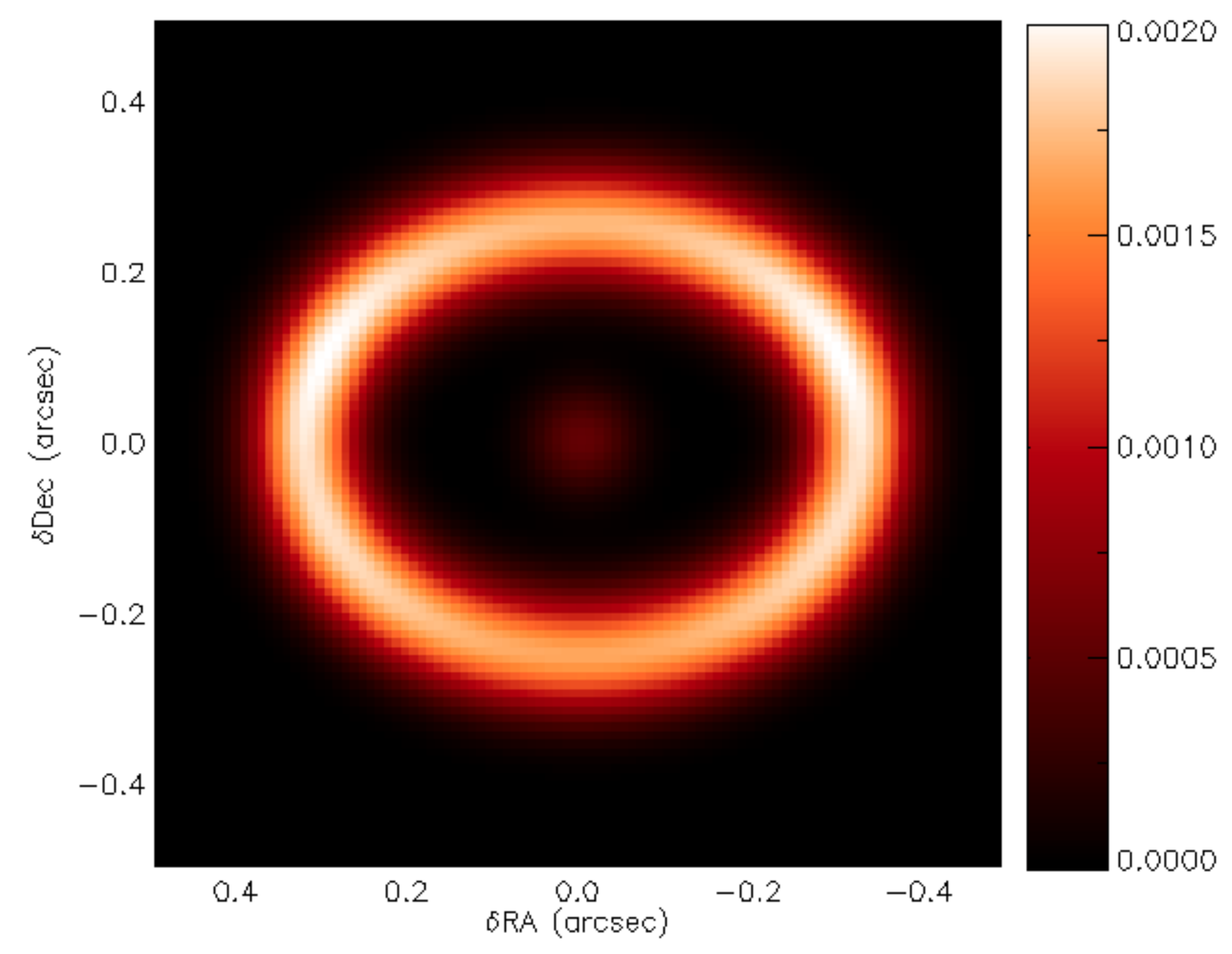}
    \end{tabular} 
   \caption{Left: 1.3~mm map in the 1~$M_\mathrm{Jup}$ case, convolved with a beam of 0.1"~$\times$~0.1". Units are in Jy/beam. Right: same for 9~$M_\mathrm{Jup}$.}
   \label{Fig:maps}
\end{figure*}

For $\alpha~=~10^{-3}$ turbulence mixing is less strong and in the range of  ($\sim$ 22-30)~AU, the pressure gradient is greater (Fig.~\ref{Fig2}-left panel), and is high enough to stop the rapid inward drift of the particles. The particles are quickly trapped at the location of the pressure maximum ($\sim$~30~AU) and stay there after several Myrs. The gap created in this case filters out all particles larger than a few $\mu m$ and the particles that are inside the gap eventually drift towards the star. The gap formed by 1~$M_\mathrm{Jup}$ mass planet with $\alpha~=~10^{-3}$ produces a pressure gradient high enough to create a concentration of dust particles at 10~AU from the location of the planet and the particles reach sizes of a few cm. 

As we note in Fig.~\ref{Fig2}- left panel, for 1~$M_\mathrm{Jup}$ and $\alpha~=~10^{-4}$, the pressure gradient is slightly higher than in the case of $\alpha~=~10^{-3}$, therefore the dust particles should be easily trapped in this case reaching even larger sizes (Eq.~\ref{eq4}). For comparison, we notice in the bottom panel of Fig.~\ref{Fig3},  that in the case without  planet and with the same disk and stellar parameters, the dust instead of being trapped, grows, fragments, mixes and drifts to the star without forming any mm-size pebbles in the outer regions of the disk $r~>~20$~AU.

The necessary size that particles have to reach to be trapped (Eq.~\ref{cond2}) is higher than the maximum sizes reached due to relative turbulent motion (Eq.~\ref{eq2}). The influence of turbulence for the dust dynamics is remarkable, if turbulence is not considered, particles  grow enough to be decoupled from the gas and they would be trapped because of the positive pressure gradient regardless of its strength. However, we conclude that the high turbulence causes particles to go through the gap and drift finally to the star, due to the fact that they reach high relative turbulent velocities  which favors fragmentation, so the mm-size particles become again micron-size grains which are not possible to trap with the strength of the pressure gradient produced with $\alpha~=~10^{-2}$.

\subsubsection{Effect of planet mass}
In the case of a 9~$M_\mathrm{Jup}$ planet, the pressure gradient is roughly similar for each of the values of the turbulence parameter $\alpha$ (see Fig.~\ref{Fig2}-right panel), hence for the dust density evolution we concentrate on the intermediate value of the viscosity, $\alpha~=~10^{-3}$. Figure~\ref{Fig4} illustrates the vertically integrated dust density distribution for the case of 9~$M_\mathrm{Jup}$ at different times of evolution. Similar to the case of 1~$M_\mathrm{Jup}$ planet and $\alpha~=~10^{-3}$, the dust particles grow and due to the positive pressure gradient between $\sim$~22 and $\sim$~48~AU (Fig.~\ref{Fig2}-right panel), the rapid inward drift is counteracted by the positive local pressure gradient and the particles are retained in a broad bump of $\sim$~20~AU width. The dust particles in the outer regions of the disk grow until mm sizes and remain there even after a few Myrs. It is important to mention that even when the pressure gradient reaches a similar value to the case of 1~$M_\mathrm{Jup}$ and $\alpha=10^{-2}$, for which there is not particle trapping at all, the turbulent relative velocities are lower and the range of positive gradient is wider, enabling dust retention  \citep{pinilla11}.

Comparing the dust evolution with $\alpha~=~10^{-3}$ for a 1~$M_\mathrm{Jup}$ (Fig.~\ref{Fig3}-bottom panel) and a 9~$M_\mathrm{Jup}$ (Fig.~\ref{Fig4}), we can conclude that the planet mass directly influences the position where the dust can be retained and that the dust bump is wider for the more massive planet. In addition, the pressure gradient created by the different planet masses also affects the size for which particles are trapped. For the same turbulence parameter $\alpha~=~10^{-3}$ and 1~$M_\mathrm{Jup}$, the maximum $a_{\mathrm{critical}}$ is equal to $\sim~0.70$~cm, while for 9~$M_\mathrm{Jup}$ the maximum $a_{\mathrm{critical}}$ is $\sim~0.47$~cm.

\subsubsection{Dust to gas ratio}
It is important to note that for the dust evolution models, we assume low values for the dust-to-gas ratio and we do not consider to have any feedback from the dust to the gas, since the gas density remains static. However,  when the dust accumulates and grows forming a pile-up,  the dust-to-gas ratio can reach values close to one. When the dust-to-gas density ratio is close to unity, the feedback from the dust to the gas is non negligible and the growth timescales can be faster than the drift timescales \citep{youdin05}, potentially leading to planetesimal formation in the absence of self-gravity. Nevertheless, these streaming instabilities are powered by the relative drift velocities between dust particles and gas, therefore in the pressure maximum where the radial drift of the dust is reduced, these instabilities are unlikely and our assumptions are still valid.

Figure~\ref{Fig_ratio} shows  the dust-to-gas ratio for each considered case of the planet mass after 1~Myr of dust evolution. The maximum value that the dust-to-gas ratio  is $\sim\{$0.76, 0.36, 0.18\} in the cases of 1, 9 and 15~$M_\mathrm{Jup}$, respectively. This variation of dust to gas ratio with the planet mass can be explained considering that as the pressure bump gets wider, the dust accumulates and grows in an extended radial region. The dust-to-gas ratio is therefore higher for lower planet mass, when the pressure gradient is higher and positive in a narrower region (see Fig~\ref{gradient}). In addition, we can notice in Fig.~\ref{Fig_ratio} that for the 15~$M_\mathrm{Jup}$ mass planet, the dust-to-gas ratio reaches very low values at the location of the planet, for which the gas drag and drift is inefficient. Therefore the remaining dust in the inner region is just the initial dust that was initially in the inner part. If the initial size of the particles is taken smaller (e.g. 0.01~$\mu m$), the sub-micron size particles would go through the gap and replenish the inner part. However, for the 1 and 9 ~$M_\mathrm{Jup}$ cases, the micron size dust from the outer region which are not trapped, can drift and be dragged by the gas and finally move through the gap.

\section{Observational predictions}  \label{sect_obs}

\subsection{Emission maps}

In this section, we present continuum maps computed at a wavelength of 1.3~mm.  We use the gas  profiles described in Sect.~\ref{sec2_2} and dust size distributions determined in Sect.~\ref{sec3_1_3} and shown in Fig.~\ref{Fig3}-middle panel and Fig.~\ref{Fig4}. We consider the same aspect ratio as in the FARGO and dust evolution simulations, and compute the volume density following:

\begin{equation}
\rho(r,z) = \frac{\Sigma(r)}{\sqrt{2\pi}h(r)} \times\exp \left( -\frac{z^2}{2 h(r)^2} \right),
  \label{eq:dens}
\end{equation}

with $\Sigma$, the dust surface density and $h(r)$, the scale height. 
After retrieving the optical constants\footnotemark{}\footnotetext{$\textrm{http://www.astro.uni-jena.de/Laboratory/Database/databases.html}$},  we compute the opacities for silicate grains, following the Mie theory.  The dust temperature is then computed self-consistently after solving the radiative transfer with the RADMC3D code \footnotemark{}\footnotetext{{\tiny $\textrm{http://www.ita.uni-heidelberg.de/$\sim$dullemond/software/radmc-3d/index.shtml}$}}.

We consider the cases of $\alpha~=~10^{-3}$, a planet with masses of 1 and 9~$M_\mathrm{Jup}$ planets and orbiting at 20~AU,  after 1~Myr. We use 40$^\circ$ and 90$^\circ$, as the disk inclination and position angle, respectively, and a distance of 140~pc. The images are convolved with a 0.10"x0.10" beam as required for ALMA for typical sizes and distances of protoplanetary disks.
As shown in Fig.~\ref{Fig:maps}, our model images are very similar to the transition disks observations in the (sub-)millimeter regime, that reveal a continuum emission as a resolved ring-like feature \citep[e.g.,][]{williams11, andrews11, isella12}. In addition, after 1~Myr of dust evolution, we estimate a spectral index of $\sim$2.7 by integrating the flux over the entire disk at wavelengths of 1~mm and 3.0~mm. This value is in agreement with recent observations \citep{ricci10}.

\begin{figure}
   \centering
   \includegraphics[width=9.0cm]{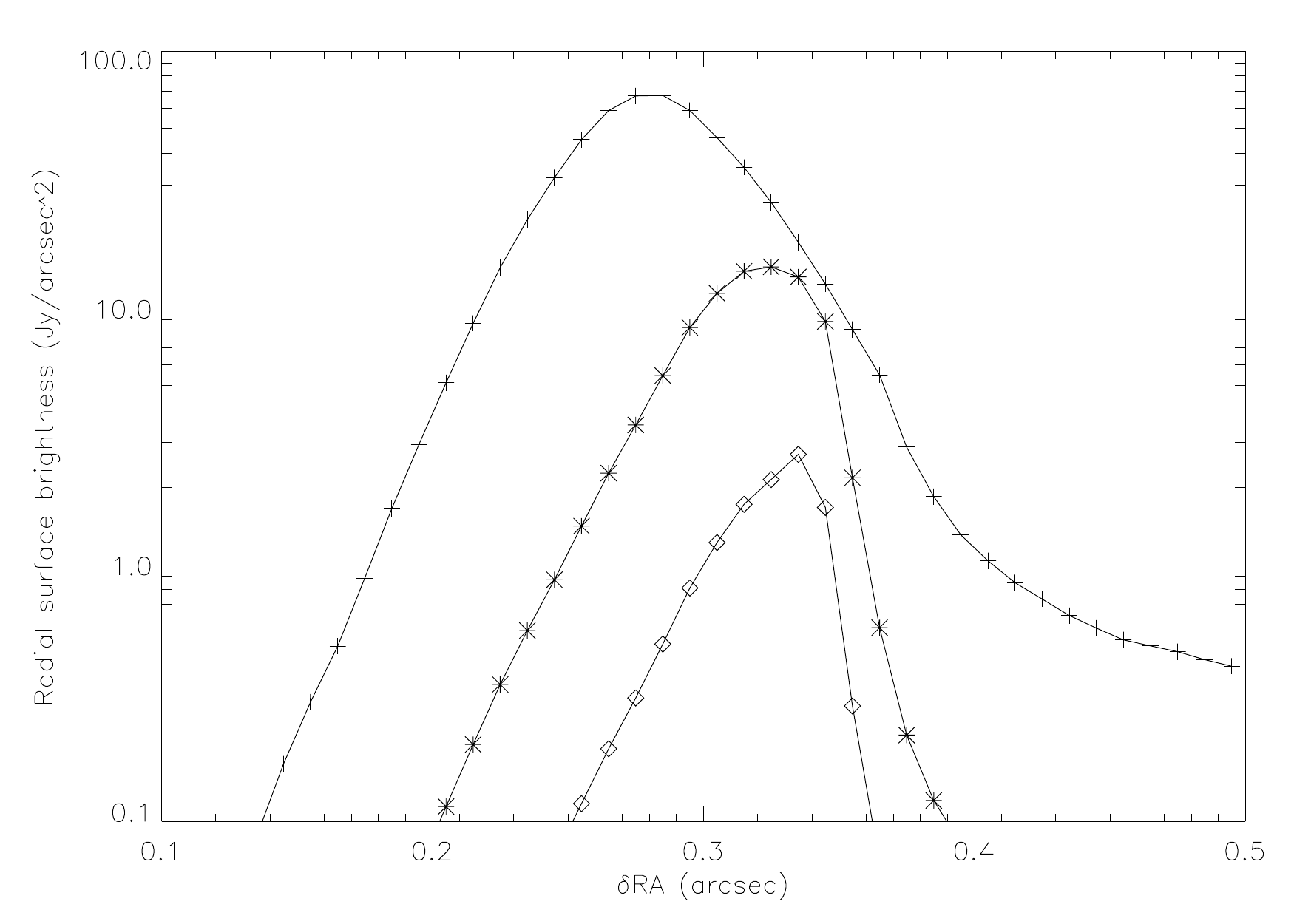}
   \caption{Radial surface brightness in the 9~$M_\mathrm{Jup}$ case, computed at 160$\mu$m (crosses), 360$\mu$m (stars) and 870$\mu$m (diamonds).}
   \label{Fig:radialcut}
\end{figure}

Due to the wide distribution of particles around the pressure bump (see, e.g., Fig.~\ref{Fig4}), the brightness distribution is expected to change with wavelength. This feature is also reproduced by our model, as shown in Fig.~\ref{Fig:radialcut}.

\subsection{The case of LkCa 15}  \label{sec4_1}
In this section, we apply and discuss our model in the context of the transition disk LkCa15, for which a wealth of observational data is published. \cite{andrews11} presented a 870~$\mu m$ image obtained at high angular resolution ($0.25^{\prime \prime}$) combining observations from the Submillimeter Array (SMA) and Plateau de Bure Interferometer (PdBI). These observations of the dust continuum emission confirms a ring-like structure at $\sim$50~AU from the star. They compared these observations with four possible models, concluding that their best fit corresponds to a gap possibly opened by a 9~$M_\mathrm{Jup}$ companion at $\sim$~16~AU in a wedge disk, with a constant disk aspect ratio of $h/r~=~0.05$ and constant viscosity $\nu=10^{-5}$ (which corresponds to $\alpha\sim10^{-3}$). In Sect.~\ref{sec3},  we demonstrated that, in the case of a 9~$M_\mathrm{Jup}$ planet and $\alpha=10^{-3}$,  the pressure maximum is located around $\sim1.4~r_p$ from the planet (with $r_p$, the planet location), which implies that the pile-up of dust (up to cm-sizes) would be located at around $\sim$~38~AU. Therefore, to have an agreement with the depletion of dust opacities until $\sim$~50~AU, a bit more massive or/and further located companion is necessary to reproduce the ring-like structure as we will discuss below. 

Although not confirmed, \cite{kraus11} claimed to have detected a $\sim~6-15~M_\mathrm{Jup}$ planet at 15.7~$\pm$~2.1~AU in LkCa15. Considering the upper limit for the mass and the location of the planet at 20~AU, the zero point for the pressure gradient would be located $\sim$~54~AU (Fig.~\ref{gradient}). Hence, dust would accumulate at $\sim~$54~AU distance from the star, in better agreement with the observed ring radius.

\begin{figure}
   \centering
      \includegraphics[width=9.0cm]{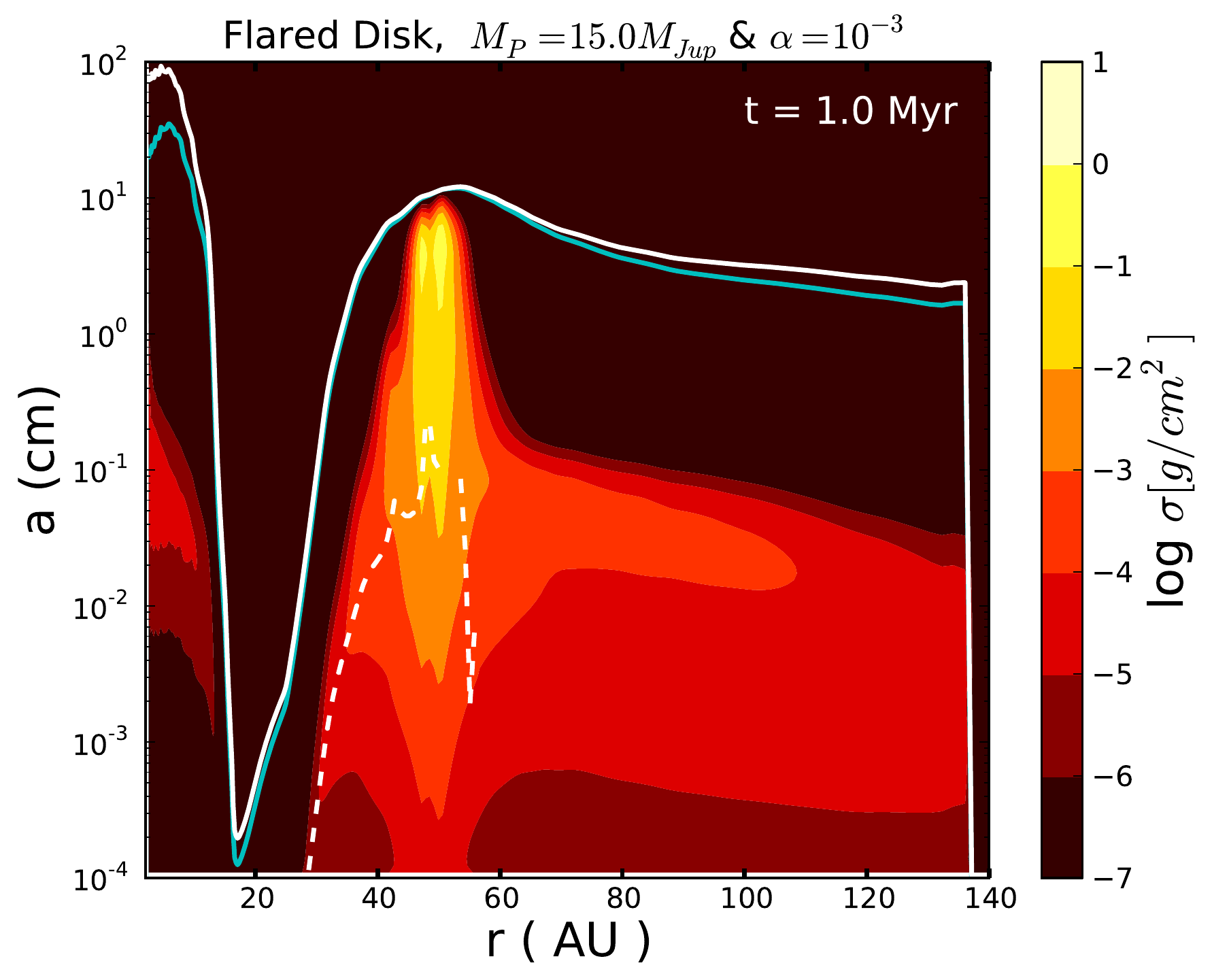}
   \includegraphics[width=9.0cm]{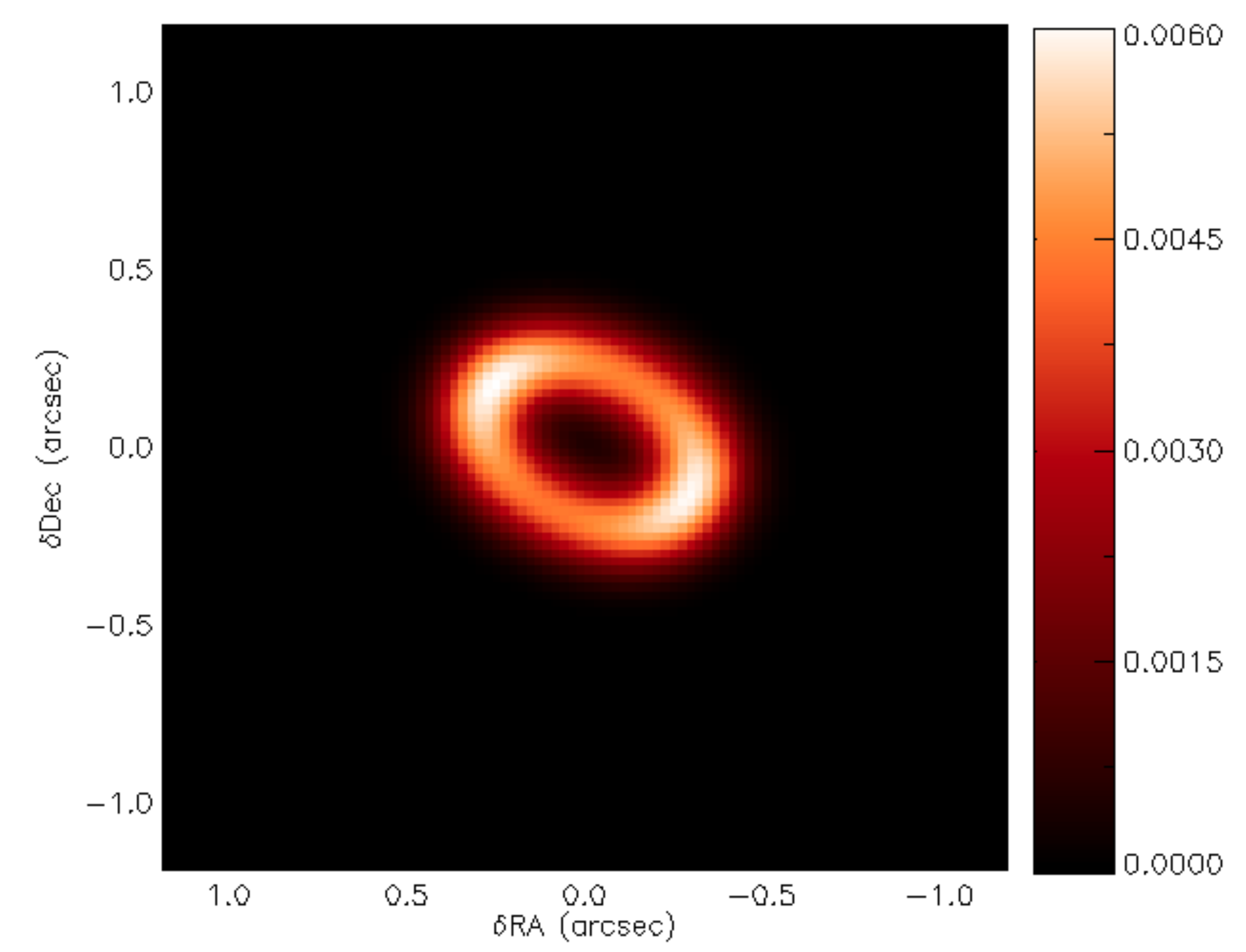}\\
   \caption{Vertically integrated dust density distribution at 1~Myr of evolution for a surface density with a gap created by 15~$M_\mathrm{Jup}$ located in a fixed orbit at 20~AU in a flared disk and $\alpha~=~10^{-3}$ (top-panel). Bottom-panel is the 1.3 mm continuum model map for LkCa15 considering the dust density distribution shown in top-panel. The map is convolved with a beam of 0.21"~$\times$~0.19", corresponding to the maximum angular resolution achieved by \citet{isella12} Units are in Jy/beam.}
   \label{Fig:maplkca15}
\end{figure}

Figure~\ref{Fig:maplkca15}-top panel shows the vertically integrated dust density distribution after 1~Myr of evolution in the case of a gap opened by a 15~$M_\mathrm{Jup}$ planet, embedded in a flared disk with $\alpha=10^{-3}$ and the same parameters given in Tables~\ref{ref:table1} and \ref{ref:table2}. Considering this dust distribution, Fig.~\ref{Fig:maplkca15}-bottom panel gives the 1.3~mm continuum map, considering a disk inclination of 51$^\circ$, a disk position angle of 64.4$^\circ$, a distance of 140~pc, and convolved with a beam of 0.21$^{\prime\prime}$$\times$ 0.19$^{\prime\prime}$ \citep{isella12}. This image shares striking similarities with the ones published, with similar inner radii for the ring and similar surface brightness. We note however, that we do not reproduce the observed width of the ring, as after 1~Myr already, most of the outer disk particles are retained in a very narrow region. Nevertheless, \cite{pinilla11} demonstrated that the presence of long-lived pressure bumps such as results from MRI can explain the presence of mm-size grains in the regions of the disk outside 50~AU. The combination of those pressure inhomogeneities and the interaction between a massive planet with the disk may explain the radial extent. 

Unlike our models with an alpha-viscosity and a massive planet embedded in the disk, \cite{isella12} model a gas density profile which increases with radius for the inner part due to the extremely rapid increase of the viscous  accretion velocity for smaller radius. 
They fit their CARMA 1.3 mm observations assuming no planet with the following gas surface density:

\begin{equation}
\Sigma(r)=\Sigma_t \left( \frac{r_t}{r}\right)^{-\gamma}\times \exp \left\{-\frac{1}{2(2-\gamma)} \left[ \left(\frac{r}{r_t}\right)^{2-\gamma}-1\right]\right\},
\label{eq8}
\end{equation}

\noindent where the viscosity is given by $\nu(r)~\propto~r^\gamma$. Their best fit model corresponds to $\gamma~=~-2.15$, $r_t~=~60$~AU and $\Sigma_t~\sim~10.37~\mathrm{g.cm}^{-2}$, for a disk extending from 0.2-160~AU. Their gas density profile suggests that the pressure gradient would be positive from the inner radius until the characteristic radius $r_t$ and smoothly becomes negative until the outer radius of the disk. This implies that the dust particles would have a positive drift velocity in the inner regions, and negative in the outer regions of the disk. They would therefore move outwards from the inner radius to $r_t$, and from the outer radius inwards to $r_t$, creating a wide concentration of dust around 60~AU. 

This complementary study, that uses a prescription of the surface density based on the viscous evolution of the disk, leads to the same conclusion as our model, i.e.,  a pressure maximum around 60 AU. In fact, the profiles are very similar in the range of radii in which the dusty ring is observed (see Fig.~1 in \citet{isella12}).

\section{Discussion} \label{sec4}
Hydrodynamical simulations of planet/disk interactions combined with dust coagulation/fragmentation models can reproduce the ring-like emission of some transition disks. 

\subsection{Dust evolution}
To compute the evolution of dust, one has to account for four important physical mechanisms: coagulation, fragmentation, radial drift and turbulent mixing. Coagulation is important for grains to grow to large sizes, as inferred from millimeter observations. Fragmentation continuously replenishes the disk in small grains.  If we do not consider fragmentation, in the region where the pressure gradient is positive, particles would accumulate, grow above the fragmentation barrier  and reach the maximum size of $\sim$2~m considered  in our models. This  would happen on  timescales as short as $\sim$~0.1~Myr, and would result in a disk only made of large particles.  As meter-size objects have St~$\gg~1$, they would therefore not be coupled with the gas anymore, move at Keplerian velocity, and would not be trapped. These grains would in addition have very low opacities, which would result in the underestimation of millimeter fluxes in the millimeter.

Radial drift is essential to distribute the grains over the disk extent. \cite{birnstiel10b} showed that if the radial drift is neglected, dust grains could acquire millimeter sizes,  but would lead to over-predictions of mm-fluxes. In our model of LkCa15, we overcome this issue and obtain flux levels similar to the ones measured by \citet{isella12}.

Finally, turbulent mixing provides the necessary relative velocities for grains to fragment, directly influencing the maximum particle size  (Eq. \ref{eq4}), and distribute the grains of various sizes all over the disk. In Sect. \ref{sec3_1_3}, we showed that the efficiency of particle trapping depends on the viscosity of the disk. For a shallow gap (see Fig \ref{Fig3}), we concluded that when the pressure gradient is not high enough, drag forces and diffusion do not allow particles to be trapped.

Our model can however still be refined. Recent collision experiments with silicates show that particles should also bounce as an intermediate  step between sticking and fragmentation \citep{guttler10}. In fact, \cite{windmark12} showed that with the insertion of cm-sized seeds, considering mass transfer processes and  bouncing, at a distance of few AU, small dust can grow to $\sim$~100 meters on timescales of 1~Myr. A similar process could happen to mm-sized particles in the outer regions of the disks. However, in the presence of ices, the need to account for bouncing effects is still highly debated \citep[e.g.][]{wada11},  and we do not consider it in this work.

\subsection{Different gap extents in the dust and the gas}
The presence of a planetary or sub-stellar companion has been suggested in many studies to explain the properties of transition disks. However, studies based on hydrodynamical simulation of gap opening in a gaseous disk have shown that a single planet could not open the large gap inferred from spatially resolved millimeter observations \citep[e.g.][]{dodson11, zhu11}. Most studies in fact consider a maximum radius for the gap of 5~Hill radii. According to the definition of the Hill radius,  the increase of  the  planet mass does not produce a significant widening of the gap, and none of these models with a single planet can reproduce the observed characteristics of transition disks. In this study, we include for the first time the modelization of dust growth and evolution, in addition to the hydrodynamical simulations. We predict grain size distributions (and therefore brightness distributions) and show that the gap edges are different in dust and gas, and that therefore the extent of the gap due to a single planet is larger than expected based on gas simulations. We demonstrate in Sect. \ref{sec3} that for a given planet, dust particles are trapped further out than the location of the edge of the gap in the gas. For example, in the specific case of a 15~$M_\mathrm{Jup}$ planet at 20~AU, the outer edge for the gas would be located at 37~AU, while the dust ring would be at 54~AU.

For $M_p~\geq~5~M_\mathrm{Jup}$, the density waves produced by the planet are still changing with time after few thousands orbits. \cite{kley06} showed that  the disk becomes eccentric with the presence of a planet with a mass $M_p>3M_\mathrm{Jup}$ (for a constant viscosity given by $\nu=10^{-5}$), which may cause a change in the equilibrium state. However, the disk precesses slowly enough that the eccentric pattern appears to be nearly stationary, and for a critical  mass of  5~$M_\mathrm{Jup}$, the eccentricity of the disk reaches a threshold. In fact, we demonstrated in  Sect.~\ref{sec3_1_2}  that for 1 and 3~$M_\mathrm{Jup}$ the dust would accumulate at 7~$r_H$ from the location of the planet $r_p$,  and at 10~$r_H$ from $r_P$ for 5, 9 and 15~$M_\mathrm{Jup}$. We attribute this difference in the locations of the zero point of the pressure gradient to the fact that the disk becomes eccentric for planets whose mass is higher than 3~$M_\mathrm{Jup}$. As a result, we have the pressure maximum further away for more massive planets, when the disk is eccentric, than for less massive planets.
 
\subsection{The inner disk}  \label{sec4_2}
Near-infrared excesses are measured in some transition disks (sometimes called pre-transition disks), and indicate the presence of dust close to the star,  that can be spatially resolved in a few cases \citep[e.g.,][]{olofsson11, benisty10, pott10}. Besides, these objects show signs of on-going accretion \citep[see e.g][]{espaillat07}. In this paper, we do not treat the inner disk because our prescription for it is not accurate. 
In order to have self-consistency between gas hydrodynamical simulations and dust evolution models, we considered for the kinematic viscosity $\nu=\alpha~c_s~h$. With this viscosity and the two considered assumptions for the gas density profiles, the accretion rate $\dot{M}~=~3~\pi~\nu~\Sigma$  is independent of the radius. With the disk and stellar parameters considered  for the two cases (Tables \ref{ref:table1} and \ref{ref:table2}), $\dot{M}$ is of the order of $\thicksim 10^{-9}~M_{\odot}~yr^{-1}$, about an order of magnitude lower than the measured values in some transitional disks ($\thicksim 10^{-8}~M_{\odot}~yr^{-1}$; \citet{calvet05, espaillat07}). The low model value implies that even considering open boundary conditions, the gas surface density, in both inner and outer regions, only slightly decreases over the 1000 orbits (Fig.~\ref{Fig1}). In addition, the insertion of a very massive planet in the disk leads to strong density waves that may lead to unrealistic effects at the inner boundary \citep[e.g.,][]{crida07b}. Therefore, our prescription for the inner disk is not accurate. 

Besides, it is important to notice that inside the ice-line the absence of ices can change our results. In fact, according to laboratory experiments, the fragmentation velocity varies with the material properties  \citep[e.g.][]{gundlach11}.  In the presence of ices, the fragmentation velocity reaches a value of $v_f \sim~1000\mathrm{cm~s}^{-1}$, the value that we consider for the outer disk. However, this threshold decreases in the inner AUs, inside the ice-line. This directly impacts the maximum particle size (Eq. \ref{eq4}). 

Finally, the radial drift in our model leads to empty the inner disk in timescales that can be shorter than the age of some transition disk with NIR excess (e.g., HD100546). Replenishment from the outer disk with small grains passing through the gap, or from planetesimal collisions \citep{krijt11} is needed.

\section{Summary and conclusions} \label{sec5}

The sub-Keplerian radial velocity of the gas in protoplanetary disks makes millimeter-size particles  in the outer regions of the disk  exposed to a rapid inward drift, implying that they migrate towards the star on timescales shorter than a million years. Therefore, in planet formation, rapid inward drift is one of the main issues with models to form planetesimals. The idea of the presence of pressure bumps in protoplanetary disks has been proposed as a solution to stop the rapid inward drift. With the presence of a massive planet in a disk, a pressure bump is  created in the outer edge of the cleared gap. Particles may experience a positive pressure gradient and as a result,  do not drift anymore and not experience the high-velocity collisions due to relative radial and azimuthal drift. Nevertheless, particles can still fragment due to turbulence motion, and the resulting micron-size particles are less easily trapped and they can finally drift inwards.

Some transition disks reveal gaps that can result from the presence of a massive planet, making them potentially interesting laboratories for studying dust growth under the favorable circumstances of having a significant pressure bump.  In this paper, we combine two-dimensional hydrodynamical simulations for the gas with one-dimensional coagulation/fragmentation dust evolution models, to study how dust evolves in a disk which gas density has been disturbed by a massive planet.

We investigate the influence of the disk geometry, turbulence and planet mass on the potential trapping of particles. The disk geometry does not have a significant effect on the gap opening process.  For a 1~$M_\mathrm{Jup}$ planet, there is an important influence of the $\alpha$ turbulence parameter on the depth of the gap and has important consequences for the trapping of particles. Unlike the case of  $\alpha=10^{-2}$, with $\alpha=10^{-3}$, the particles are trapped at a distance of $\sim~0.5~r_p$ from the planet position $r_p$.  While the gas gap has a radius of $\sim~0.35~r_p$ (or 5 Hill radii), the dust is retained at a further distance. Without considering turbulence mixing for the dust dynamics would produce trapping of particles for certain sizes regardless of $\alpha$ value provided that the pressure gradient is positive. In the case  of 9~$M_\mathrm{Jup}$ planet, the gap depth is independent of the turbulence and as a consequence the pressure gradient behaves similarly for all  $\alpha$ values.  The dust particles are trapped at a distance of 1.4~$r_p$ from the planet orbit $r_p$. The planet mass strongly influences the location where the dust is retained and the width of the dust bump.

Observations at millimeter wavelengths reveal some transition disks with very wide gaps.  We show that the location where the dust  piles up does not coincide with the gap outer edge in the gaseous disk. This mismatch suggests that multiple planets may not necessarily be needed to account for the observed large opacity holes. We reproduce the main observed features of the disk around LkCa15, in which a companion was recently detected. Reproducing asymmetric features in the disk, such as the ones found in HD135344B \citep{brown09} is subject of future work, as they will be well detectable with ALMA.

\begin{acknowledgements}
The authors are very grateful to C.P.~Dullemond for insightful suggestions, and for making RADMC3D available. We acknowledge C.~Dominik and A.~Natta for their enthusiasm and comments. We also thank A.~Isella and S.~Ataiee for useful discussions. P. Pinilla acknowledges the CPU time for running simulations in bwGRiD, member of the German D-Grid initiative, funded by the Ministry for Education and Research (Bundesministerium f\"ur Bildung und Forschung) and the Ministry for Science, Research and Arts Baden-Wuerttemberg (Ministerium f\"ur Wissenschaft, Forschung und Kunst Baden-W\"urttemberg). 
\end{acknowledgements}

\bibliographystyle{aa}
\bibliography{transitiondisks_pinilla}

\end{document}